\documentclass{article}

\usepackage[english]{babel}

\usepackage[letterpaper,top=2cm,bottom=2cm,left=3cm,right=3cm,marginparwidth=1.75cm]{geometry}

\usepackage{amsmath}
\usepackage{amsfonts}
\usepackage{graphicx}
\usepackage{makecell}
\usepackage{lingmacros}
\usepackage{tree-dvips}
\usepackage{graphicx}
\usepackage{caption}
\usepackage{float}
\usepackage{amsmath}
\usepackage{amssymb}
\usepackage{amsthm}

\usepackage{tikz}
\usetikzlibrary{shapes,positioning}
\usepackage{mathtools}
\usepackage[super]{nth}
\usepackage{array}
\usepackage{tabularx}
\usepackage{lingmacros}
\usepackage{tree-dvips}
\usepackage{graphicx}
\usepackage{algorithm}
\usepackage{algpseudocode}
\usepackage{multirow}
\usepackage{makecell}
\usepackage{array}
\usepackage{titlesec}
\usepackage{rotating}
\usepackage{dirtytalk}
\usepackage{lipsum}
\usepackage{titletoc}%
\usepackage{graphicx}
\usepackage{capt-of}

\usepackage[colorlinks=true, allcolors=blue]{hyperref}

\title{Hedge Fund Portfolio Construction Using PolyModel Theory and iTransformer}
\author{
Siqiao Zhao\thanks{Corresponding author.}  \\
{\tt \small alysia.zhao@gmail.com}
\and
Zhikang Dong \\
Stony Brook University\\
{\tt \small zhikang.dong.1@stonybrook.edu}
\and
Zeyu Cao \\
{\tt \small josephcao891011@gmail.com}
\and
Raphael Douady \\
University of Paris I: Pantheon-Sorbonne\\
{\tt \small rdouady@gmail.com}
}

\date{} 

\begin{document}
\maketitle


\section{Introduction}
Portfolio construction remains a central topic in quantitative finance research. Beginning with the Capital Asset Pricing Model (CAPM) \cite{fama2004capital}, the theory of portfolio construction has continuously evolved, incorporating a range of new techniques and theories over time. Data-driven methods, particularly in fields like computer vision \cite{li2023blip,dong2024mapping,radford2021learning,kim2023face,dan2024multiple,dan2024evaluation,zhang2023mm,zhang2023self,dong2024mamba,fu2024clipscope,dong2025every}, natural language \cite{lyu2023attention,yang2022reinforcement}, time series analysis \cite{sun2023manifold,lu2024cats}, biomedical research \cite{tang2023srda,kumar2022leveraging,dong2021detection,wei2023research} audio processing \cite{gong2021ast,dong2024musechat, liu2024tackling}, content moderation \cite{xin2024let}, statistics \cite{chencalibrate,xu2024variance} and science \cite{abramson2024accurate,dong2023cp,dong2024cp,lin2024open,fudong:iccv23:mmst_vit}, have shown significant advancements. In recent years, those techniques have notably impacted quantitative finance \cite{papanicolaou2023optimal,papanicolaou2023deep}, from predicting asset prices \cite{kumar2022systematic} to hedging risks in derivatives \cite{buehler2019deep}.

However, when constructing portfolios, a key problem is that a lot of financial time series data
are sparse, making it challenging to apply machine learning methods. Polymodel
theory can solve this issue and demonstrate superiority in portfolio construction from
various aspects. To implement the PolyModel theory for constructing a hedge fund
portfolio, we begin by identifying an asset pool, utilizing over 10,000 hedge funds for
the past 29 years’ data. PolyModel theory also involves choosing a wide-ranging set of
risk factors, which includes various financial indices, currencies and commodity prices.
This comprehensive selection mirrors the complexities of the real-world environment.
Leveraging on the PolyModel theory, we create quantitative measures such as Longterm Alpha, Long-term Ratio, and SVaR. We also use more classical measures like the
Sharpe ratio or Morningstar’s MRAR. To enhance the performance of the constructed
portfolio, we also employ the latest deep learning techniques (iTransformer) to
capture the upward trend, while efficiently controlling the downside, using all the
features. The iTransformer model is specifically designed to address the challenges in
high-dimensional time series forecasting and could largely improves our strategies.
More precisely, our strategies achieve better Sharpe ratio and annualized return. The
above process enables us to create multiple portfolio strategies aiming for high
returns and low risks when compared to various benchmarks. The integration of
PolyModel theory with machine learning methods facilitates a nuanced and precise
understanding of hedge fund returns. This amalgamation enables us to overcome
challenges related to hedge fund data, offering a more robust methodology for
analyzing hedge fund performance and guiding investment decisions. This is a very
meaningful attempt to combine fundamental statistical analysis with latest machine
learning techniques.

\section{PolyModel Theory}
The origin of the idea of PolyModel theory and its mathematical foundations can be dated back to \cite{cherny2010measuring} and \cite{coste2010stressvar}. Since PolyModel theory is more a framework rather than a single statistical analysis tool, after its first introduction, quite a few extensions and applications have been proposed and studied. For a nice overview of more applications and the history of this theory, one can check \cite{douady2019managing} while for more concise mathematical description and its implementation, one can consult \cite{barrau2022artificial} and \cite{sq}.

Before we step into the mathematical descriptions, let's first discuss the core idea and intuition behind PolyModel theory to get a better understand of it.

The core idea of PolyModel theory is to combine a large enough collection of valid description of one aspect of the same target or reality in order to get a as close as possible fully understanding of the target's nature. In financial industry, the target is usually the return of some asset in which one wants to invest.

If we image that the target is alive, like an animal, then PolyModel theory can be regarded as a methodology to observe how this animal reacts to the outside environment, especially, to each single environment factor. If we can capture and understand all its reactions, then we can fully characterize this animal. This idea is, surprisingly, similar to a Python terminology called "Duck Typing": "when an object quacks like a duck, swims like a duck, eats like a duck or simply acts like a duck, that object is a duck." Though coming from very different fields, the two ideas introduced above can both be viewed as an variant of Phenomenology  \cite{barrau2022artificial}: "Literally, phenomenology is the study of 'phenomena': appearances of things, or things as they appear in our experience, or the ways we experience things, thus the meanings things have in our experience."

After the high-level description of PolyModel theory, we now turn back to its mathematical descriptions and how to construct features with strong description or prediction power.

\subsection{Mathematical formulation and model estimation}

\subsubsection{Model description and estimation}

 There are two fundamental components in PolyModel theory:
 \begin{itemize}
     \item A pool of target assets $\{Y_i\}_{i \in I}$ which are the components of the portfolios one want to construct.
     \item A very large pool of risk factors $\{X_j\}_{j \in J}$ which form a proxy of the real-world financial environment.
 \end{itemize}

 The mathematical description of the PolyModel theory can be formulated as follows:

 For every target $Y_i$, $i \in I$, there is a collection of (relatively simple) regression models:
 \begin{equation}
     \begin{cases}
        Y_i = \Phi_{i1}(X_1) + \epsilon_1\\
        Y_i = \Phi_{i2}(X_2) + \epsilon_2 \\
        ... ... \\
        Y_i = \Phi_{in}(X_n) + \epsilon_n
    \end{cases}
 \end{equation}
where
\begin{itemize}
    \item $n$ is the number of the risk factors.
    \item $\Phi_{ij}$ is assumed to capture the major relationship between independent variable $X_j$ and dependent variable $Y_i$; in practice, it is usually a polynomial of some low degree.
    \item $\epsilon_{j}$ is the noise term in the regression model with zero mean; usually it is assumed to be normal distribution but does not have to be.
\end{itemize}

In practice, we usually assume that
\begin{equation}
    \Phi_{ij}(x) = \Sigma_{k = 0}^{4} \beta_{ij}^k H_{k}(x),
\end{equation}

where $H_{k}(x)$ is the Hermitian polynomial of degree $k$. Based on authors' practical experience, a polynomial of degree of 4 is flexible enough to capture nonlinear but essential relation between target and risk factor while usually suffer bearable overfitting.

For each target and risk factor pair $(Y_i, X_j)$, assume that we have their observations: $Y_i$ and $X_j$ for time $t = 1,2, ..., T$, then we can write each regression model from (1) into matrix format

\begin{equation}
    \overrightarrow{Y_i} = \boldsymbol{H}_j^{T} \overrightarrow{\beta_{ij}} + \overrightarrow{\epsilon_{ij}},
\end{equation}

where
\begin{itemize}
    \item $\overrightarrow{Y_i}$ denotes the vector of the target time series such of return of hedge fund
    \begin{center}
        $\begin{bmatrix}
           Y_i(t_1) \\
           Y_i(t_2) \\
           \vdots \\
           Y_i(t_T)
         \end{bmatrix}$.
    \end{center}  

    \item $\boldsymbol{H_j}$ denotes the following matrix of the risk factor $X_i$
    \begin{center}
        $\begin{bmatrix}
           H_0(X_j(t_1)), \  H_0(X_j(t_2)), \  H_0(X_j(t_3)), ..., \  H_0(X_j(t_T)) \\
           H_1(X_j(t_1)), \  H_1(X_j(t_2)), \  H_1(X_j(t_3)), ..., \  H_1(X_j(t_T)) \\
           \dots\\
           H_4(X_j(t_1)), \  H_4(X_j(t_2)), \  H_4(X_j(t_3)), ..., \  H_4(X_j(t_T))
         \end{bmatrix}$.
    \end{center}
    which is a $5 \times T$ matrix, where $H_k(x)$ is the Hermitian polynomial of degree $k$.
    
    \item $\overrightarrow{\epsilon_{ij}}$ denotes the regression error vector
    \begin{center}
        $\begin{bmatrix}
           \epsilon_{ij}(t_1) \\
           \epsilon_{ij}(t_2) \\
           \vdots \\
           \epsilon_{ij}(t_T)
         \end{bmatrix}$.
    \end{center}

    \item $\overrightarrow{\beta_{ij}}$ is the coefficient vector of length $5$
    \begin{center}
        $\begin{bmatrix}
           \beta_{ij}^{0} \\
           \beta_{ij}^{1} \\
           \vdots \\
           \beta_{ij}^{4}
         \end{bmatrix}$.
    \end{center}.
    
\end{itemize}

Now let's briefly discuss how to estimate the coefficients. From the model description above, we can see that PolyModel theory technically belongs to the realm of statistical regression models, thus, all the common well-established parameter estimation methods can be applied to it. From a practical point of view, we choose to use the Ridge regression \cite{hastie2009elements}

\begin{equation}
 \widehat{\overrightarrow{\beta}}_{ij, \lambda} := arg \  min_{\{\overrightarrow{\beta_{ij}} \in R^5\}} 
    [(\overrightarrow{Y_i} - \boldsymbol{H}_j^{T} \overrightarrow{\beta_{ij}})^T (\overrightarrow{Y_i} - \boldsymbol{H}_j^{T} \overrightarrow{\beta_{ij}}) + \lambda ||\overrightarrow{\beta_{ij}}||^2,
\end{equation}

We can see that the fitted coefficients are functions of the hyper-parameter $\lambda$; to determine the optimal value for each simple regression, one can apply any state-of-art hyper-parameter tuning trick such as grid search plus cross-validation. However, we would like to point out that in PolyModel theory, we need to deal with a huge amount of risk factors, and our polynomial in the regression equation is only of degree 5, thus, our major concern for using ridge regression is to make the matrix $\bold{H}_j \bold{H}_j^{T} + \lambda I_{5 \times 5}$ invertible, thus, we usually choose a relatively small number as the value of $\lambda$ for all the target time series and risk factor pairs.

\subsection{Feature Importance and Construction}

One of the major goals of PolyModel theory is to find a set of risk factors which are most important to the target time series after fitting hundreds of simple regressions. In this section, we will first discuss the fundamental statistical quantities based on fitting the numerous simple regressions, then we will use them as building blocks to construct the features which will be used by the machine learning algorithms.

\subsubsection{Fundamental statistical quantities}

\begin{enumerate}
    \item $R^2$ and adjusted $R^2$
    
    As PolyModel is a collection of simple regression models, then it is quite natural to talk about $R^2$ for every simple regression model.

    $R^2$, also known as coefficient of determination, is one of the most common criteria to check the fitting goodness of a regression model. It is defined as follows:

    \begin{equation}
    R^2 := \frac{ESS}{TSS} = 1 - \frac{RSS}{TSS},
    \end{equation}
    where, if we denote $\bold{H}_j^{T} \widehat{\overrightarrow{\beta}}_{ij}$ by $\widehat{\overrightarrow{Y_i}}$, and denote the vector of average of entries of $\overrightarrow{Y_i}$ with the same length by $\overline{Y_i}$, then
    \begin{itemize}
        \item ESS is the explained sum of squares which is $(\widehat{\overrightarrow{Y_i}} - \overline{Y_i})^T (\widehat{\overrightarrow{Y_i}} - \overline{Y_i})$.
        \item RSS is the residual sum of squares which is $(\overrightarrow{Y_i} - \widehat{\overrightarrow{Y_i}})^T (\overrightarrow{Y_i} - \widehat{\overrightarrow{Y_i}})$.
        \item TSS is the total sum of squares which is $(\overrightarrow{Y_i} - \overline{Y_i})^T (\overrightarrow{Y_i} - \overline{Y_i})$.
    \end{itemize}

    Moreover, it is a well-known fact in regression theory that TSS = RSS + ESS.
    
    $R^2$ measures how much total uncertainty is explained by the fitted model based on the observed data, thus, the higher $R^2$ is, the better the model should be. However, this statistic does not take the number of model complexity into consideration, thus, a high $R^2$ may also indicates overfitting and usually this is the case (for instance, in a one dimension problem given general $n$ data points, there is usually a degree $n+1$ polynomial which can pass through every one of them). Various modifications have been introduced, one very direct generalization is the adjusted-$R^2$: $1 - \frac{\frac{RSS}{(n-p)}}{\frac{TSS}{(n-1)}}$ where $n$ is the number of observations and $p$ is the number of coefficients in the regression model.

    \item Target Shuffling and $P$-Value Score

    To avoid fake strong relationship between target and risk factors, we apply target shuffling which is particular useful to identify "cause-and-effect" relationship. By shuffling the the targets, we have the chance to determine if the relationship fitted by the regression model is significant enough by checking the probability of the $R^2$ we have seen based on the observations.

    The procedure can be summarized as follows:

    \begin{itemize}
        \item Do random shuffles on the target time series observations many times, say N times. For each $X_j$, let we assume that there are T data points $\{(Y_i(t_k),X_j(t_k)\}^T_{k=1}$. We fix the order of $X_j(t_k)$, and we do N times of random shuffle of $Y_i(t_k)$. In this way, we try to break any relation from the original data set and create any possible relations between the target and risk factor.

        \item For each newly ordered target observations $\{(Y'_i(t_k),X_j(t_k)\}^T_{k=1}$, we can fit a simple regression model and calculate the $R^2$. Then we get 

        \begin{center}
        $R^2_{shuffle} = \{R^2_{(1)},R^2_{(2)}, \cdots, R^2_{(N)}\}$.
        \end{center}

        Thus, we have a population of $R^2$ based on above procedures.

        \item Evaluate the significance of the $R^2$ calculated from the original data, for instance, we can calculate the p-value of it based on the $R^2$ population from last step. Here we assume that our original $R^2$ for target asset $Y_i$ and risk factor $X_j$ is denoted as $R^2_{ij}$. Then, we could define 

        \begin{center}
        $p_{ij} = P(R^2 > R^2_{ij})$.
        \end{center}
        
        \item We compute $-log(p_{ij})$ and call it $P$-Value Score of target asset $Y_i$ and risk factor $X_j$ which indicates the importance of the risk factor $X_j$ to the target asset time series $Y_i$.
    \end{itemize}

    The higher the $P$-Value Score is, the more important the risk factor is. As we  also need to take different regimes over the time into the picture, at each time stamp, we only look at the past 3 years' return data, and thus, we can have a dynamic $P$-Value Score series for each target asset $Y_i$ and risk factor $X_j$ pair.

\end{enumerate}

\subsubsection{Feature construction}

Now we are ready to construct the features based on the statistical quantities introduced above and the data themselves. We will briefly discuss how to construct them and their meanings. More detials can be found in \cite{sq}.

\begin{enumerate}
    \item Sharpe Ratio

    It is one of the most common statistical metric to estimate the performance of a portfolio. Roughly speaking, it is the ration between the portfolio return and its volatility, thus, usually is regarded as a measure of the ratio between reward and risk.

    Assume $R$ represents the return of the target portfolio, $R_{f}$ represents the return of the benchmark financial time series, for instance, RFR. Then Sharpe Ratio is defined as
    \begin{center}
        Sharpe Ratio := $\frac{E(R - R_{f})}{\sqrt{var(R - R_{f})}}$.
    \end{center}

    In practice, one may also ignore the benchmark if it is very small or static. Notice that Sharpe Ratio is a feature that is only dependent on target portfolio itself.
    
    \item Morningstar Risk-adjusted Return (MRaR)

    This is another feature mostly dependent on the target portfolio itself. Given the target portfolio (e.g. hedge fund return $Y_i$), denote its return at time $t$ as $r_t$; denote the return of benchmark at time $t$ as $r_f$, the MRaR over $n$ months is defined as follows \cite{MRaR_formula}

    \begin{center}
        $MRaR = (\frac{1}{n}\Sigma_{i=1}^{n}(1 + r_{Gt})^{-\gamma})^{-\frac{n}{\gamma}} - 1$,
    \end{center}
    
    \begin{center}
        $r_{Gt} = (\frac{1+r_t}{1+r_f}) - 1$,
    \end{center}
    
    where $n$ is the total number of months in calculation period; $r_{Gt}$ is the geometric excess return at month t; $\gamma$ is the risk aversion parameter, and $Morningstar^{TM}$ uses 2. Investors can adjust the value of $\gamma$ according to their own risk flavors. 

    As mentioned in \cite{MRaR}, the main assumption is that investors are rational and willing to give up a small portion of their expected return to achieve a better certainty. This is metric is similar to Sharpe ratio but has more advantages. More discussions on its advantages can be found in \cite{MRaR_explain}.
    
    \item StressVaR (SVaR)

    SVaR can be regarded as a good alternative risk measure instead of VaR, in fact, it can be regarded as a factor model-based VaR. However, its strength resides in the modeling of nonlinearities and the capability to analyze a very large number of potential risk factors\cite{coste2009stressvar}.

    There are three major steps in the estimation of StressVaR of a hedge fund $Y_i$.

    \begin{enumerate}
        \item Most relevant risk factors selection: for each risk factor $X_j$, we can calculate the $P$-Value Score of it with respect to $Y_i$. Recall Section 2.5.2, this score can indicate the explanation power of risk factor $X_j$, and the application of target shuffling improves the ability of our model in preventing discovering non-casual relations. Once a threshold of $P$-Value Score is set, we can claim that all the risk factors $X_j$ whose $P$-Value Score is above the threshold are the most relevant risk factors, and denote the whole set of them as $\Gamma_{i}$.
        
        \item Estimation of the Maximum Loss of $Y_i$:
        For every risk factor $X_j \in \Gamma_i$, using the fitted polynomial for the pair $(Y_i, X_j)$, we can predict the return of $Y_i$ for all risk factor returns from $1st$ to $99th$ quantiles of the risk factor distributions. In particular, we are interested in the potential loss of $Y_i$ corresponding to $\alpha \% = 98 \%$ of the factor returns. Once this is estimated for one factor $X_j$, we can define $SVaR_{i,j}$ for the pair $(Y_i, X_j)$ as follows:
    
        \begin{equation*}
            SVaR_{i,j} := \sqrt{\hat{Y}_{i,j, max}^2 + \sigma(Y_i)^2 \cdot (1 - R^2) \cdot \xi^2}
        \end{equation*}
        where
        \begin{itemize}
            \item $\hat{Y}_{i,j, max}$ is the maximum potential loss corresponding to $\alpha$ quantile of risk factor $X_j$.
            \item $\sigma(Y_i)^2 \cdot (1 - R^2)$ is unexplained variance under the ordinary least square setting which can be estimated by the following unbiased estimator if penalty terms are added to the regression models
            \begin{center}
                $\frac{\Sigma (Y_i - \hat{Y}_i)^2}{n-p}$,
            \end{center}
            where $p$ is the degree of freedom of the regression model.
    
            \item $\xi = \varphi^{-1}(\alpha) \approx 2.33$ where $\varphi $ is the cumulative distribution function (cdf) of standard normal distribution.
    
        \end{itemize}
        
        \item Calculation of StressVaR: The definition of StressVaR of $Y_i$ is
        \begin{center}
            $SVaR_i := max_{j \in \Gamma_i} SVaR_{ij} $.
        \end{center}
        
    \end{enumerate}
    
    \item Long-term alpha (LTA)

    For the given hedge fund and risk factor pair $(Y_i, X_j)$, assume we already fitted the regression polynomial $\Phi_{ij}(x)$. Assume that $\theta_{j, q}$ represents the $q$-quantile of the empirical distribution of $X_j$ where $q = 1\%, \  16\%, \  50\%, \  84\%, \  99\%$. They are calculated using the very long history of the factor. The extremes $1\%$ and $99\%$ are computed by fitting a Pareto distribution on the tails.

    Then we define
    \begin{center}
        $LTA(Y_i, X_j) := \Sigma_{q = 1 \%}^{99 \%} w_q \Phi_{ij}(\theta_{j, q})$,
    \end{center}
    subject to $E(X_j) = \Sigma_{q = 1 \%}^{99 \%} w_q \theta_{j, q}$, where $w_q$ correspond to Lagrange method of interpolating an integral and are hyper-parameters.

    The global LTA (long-term average) is the median of the factor expectations for selected factors. $LTA_i$ for $Y_i$ is defined as the $50th$ quantile among all the LTA($Y_i$, $X_j$) values, where $X_j \in \Gamma_i$ represents the selected ones.

    \item Long-term ratio (LTR)

    Once we get the $LTA_i$ and $SVaR_i$ for $Y_i$, $LTR_i$ is simply defined as

    \begin{center}
        $LTR_i := \frac{LTA_i}{SVaR_i}$.
    \end{center}
    
    \item Long-term stability (LTS)

    For fund $Y_i$, $LTS_i := LTA_i - \kappa \cdot SVaR_i$ where $\kappa$ is a hyper-parameter whose value is set to $5\%$.
\end{enumerate}

Besides the features constructed above, we also include some more standard features for our financial time series research: asset under management (AUM) of each hedge fund, volume of each hedge fund, and historical returns for each hedge fund and risk factor. All of them will be used as input features when applying machine learning techniques below. 

\section{Methodology}

Given the carefully chosen risk factor pool and the set of hedge funds to invest, we first apply PolyModel theory to construct the features introduced in the previous section. Notice that these features can be regarded as a dynamical encoding of the hedge funds' returns and their interactions with the whole financial environment.

We then will apply modern machine learning algorithms to predict the performance of each hedge fund. We particularly choose to apply transformer techniques in our prediction due to its string performance in time series related forecasting researches during recent years \cite{wen2022transformers}. Moreover, we will apply one of its latest variants called inverted transformer in our study.

In the rest of this section, we first introduce inverted transformer, then discuss how to apply it to our hedge fund performance prediction task in details.

\subsection{Inverted Transformers (iTransformer)}
Inverted Transformers (iTransformer) \cite{liu2023itransformer} is designed for multivariate time series forecasting. We combine this method with PolyModel theory to generate effective portfolio construction. Suppose we extract $N$ features with $T$ timesteps, denoted as $\mathbf{X}=\left\{\mathbf{x}_1, \ldots, \mathbf{x}_T\right\} \in \mathbb{R}^{T \times N}$. Based on those historical observations, we can forecast the future $S$ time steps target $\mathbf{Y}=\left\{\mathbf{x}_{T+1}, \ldots, \mathbf{x}_{T+S}\right\} \in \mathbb{R}^{S \times N}$. Instead of regarding multivariate features of the same time step as a temporal token, the iTransformer tokenize the whole time series input of each feature as the token, which focus on representation learning and correlation measurement of multivariate time series. 

\begin{equation}
\mathbf{h}=\operatorname{Embedding}\left(\mathbf{X}\right),
\end{equation}
where $\mathbf{h}=\left\{\mathbf{h}_{1}, \ldots, \mathbf{h}_{N}\right\} \in \mathbb{R}^{N \times D}$. We use multi-layer perceptron (MLP) to project raw time series data into $D$-dimensional latent space. \cite{liu2023itransformer} shows that the temporal information has been processed by MLP, the position embedding in original Transformer \cite{vaswani2017attention} is not necessary anymore.

We apply Layer normalization (LN) \cite{ba2016layer} to token $\mathbf{h}$ across time steps. Unlike the common Transformer frameworks, which apply LN across different features, iTransformer \cite{liu2023itransformer} normalizes each feature token to a standard Gaussian distribution, which helps keep patterns in each feature. \cite{kim2021reversible,liu2022non} also prove that this technique are helpful in solving non-stationary time series problem.
\begin{equation}
\mathbf{H}=\frac{\mathbf{h}_n-\operatorname{Mean}\left(\mathbf{h}_n\right)}{\sqrt{\operatorname{Var}\left(\mathbf{h}_n\right)}}, n=1, \ldots, N.
\end{equation}

The original Transformer \cite{vaswani2017attention} uses the attention mechanism to process temporal information for encoded features. The iTransformer \cite{liu2023itransformer} uses this attention mechanism to model feature correlations since each token represents the whole time series data of a feature. Suppose there are linear projections $\mathbf{W}_Q \in \mathbb{R}^{D \times d_k}, \mathbf{W}_K \in \mathbb{R}^{D \times d_k} \text{ and } \mathbf{W}_V \in \mathbb{R}^{D \times d_k}$. We can obtain query, key and value matrices as $\mathbf{Q} = \mathbf{H}\mathbf{W}_Q$, $\mathbf{K} = \mathbf{H}\mathbf{W}_K$ and $\mathbf{V} = \mathbf{H}\mathbf{W}_V$. Then, the self-attention mechanism is computed as

\begin{equation}
\operatorname{Attention}(\mathbf{Q}, \mathbf{K}, \mathbf{V})=\operatorname{softmax}\left(\frac{\mathbf{Q} \mathbf{K}^T}{\sqrt{d_k}}\right) \mathbf{V}.
\end{equation}

Traditional transformer models typically utilize temporal tokens, analyzing all features at a single timestamp, which can limit their ability to effectively learn dependencies. One approach to address this limitation involves patching, where data points along the time axis are grouped prior to tokenization and embedding. However, this method may suffer from insufficiently large grouping ranges, failing to capture all necessary dependencies. In contrast, the iTransformer adopts an innovative approach by viewing the time series from an inverted perspective. This allows it to consider the same feature across multiple timestamps, significantly enhancing its capacity to discern dependencies and multivariate correlations. This distinct capability positions the iTransformer as a superior alternative in scenarios demanding nuanced temporal analysis.





\subsection{Hedge fund performance prediction}

We apply iTransformer algorithm directly in our research. The input features are those described in section 2.2.2. Regarding the output, for each target hedge fund, we predict the probability of the trend rather than the value of its return, in particular, we assume that there are three status of the return trend: up, down and unchanged (we set a prior threshold for the hedge fund return. If the absolute value of the return is smaller than the threshold, we define its status as unchanged. Otherwise, the status is up if the return is positive and the status is down if the return is negative).

We apply the implementation of iTransformer from \cite{liu2023itransformer} in a straight forward manner where interested readers can find all the technical details. Thus, rather than more discussions on iTransformer, we will discuss why we choose the trend rather than the value of hedge fund returns as our prediction output.

As already pointed out in some recent research such as \cite{sq_rio_2023}, \cite{vuletic2024fin}, it is more useful to correctly classify the trend of returns rather than to provide a predicted result which is close to the real return. For instance, one has a portfolio and can predict its return as close as the realized one but with an opposite sign, this may cause a significant negative impact on one's pnl and is not favored. 

Moreover, our target assets are hedge funds whose returns usually have very large magnitude, thus, once we can predict the return status correctly and select those hedge funds whose next returns are positive, we will have a good chance to achieve a reasonably high total return. On the other side, PolyModel theory is quite good at identifying risk factors which may cause large drops of the target assets. Thus, the combination of these two theories can give us a better chance to create a portfolio with large positive return and small drawdown.

\section{Portfolio Construction}

Based on the theories and methodologies introduced in previous sections, we are ready to construct our portfolio. We rebalance our portfolio monthly. Before the end of each month, we apply iTransformer to predict the probability on whether the return of hedge fund $Y_i$ for the next month is positive which is denoted as $p_i$. We select the top $50\%$ hedge funds with the largest probabilities of having a positive return for the next month. We keep those hedge funds which are currently held in our portfolio if they are selected, and sell the in-selected ones in our hands. The collected cash are reinvested evenly to buy the rest selected hedge funds which are not in current portfolio. We call this strategy simple average portfolio (SA). A second proposed strategy, which is denoted as weighted average portfolio (WA), is almost identical to SA except that the weights of the selected fund in the portfolio are based on the their AUM.

\section{Experiments and Results}

In this section, we will give an overview of the data used for our study, the benchmarks to compare with and the performance of our portfolio. The same set of data and benchmarks are also used in \cite{sq_rio_2023}.

\subsection{Data description}

As mentioned in the introduction of PolyModel theory, there are two datasets: risk factors and target hedge funds. The data sets cover a long period from April 1994 to May 2023. These data will be used to construct the features introduced in section 2.2.2, and the set of hedge fund will be used to construct the portfolio. Below let's look at the snapshots of some of the representatives of these two data sets.

 Regarding risk factors, our study incorporates an extensive universe comprising hundreds of risk factors from different domains, including equities, coupons, bonds, industrial indexes, and more. We list some of the risk factors:

\begin{center}
\begin{tabular}{||c c ||} 
 \hline
 Label & Code \\ [1ex] 
 \hline\hline
\makecell{T-Bil} & {INGOVS USAB} \\
 \hline
 \makecell{SWAP 1Y Zone USA In\\USD DIRECT VAR-LOG} & {INMIDR USAB}  \\
 \hline
 \makecell{American Century Zero Coupon\\2020 Inv (BTTTX) 1989} & {BTTTX}  \\
 \hline
 \makecell{COMMODITY GOLD Zone USA\\In USD DIRECT VAR-LOG} & {COGOLD USAD} \\ 
 \hline
 \makecell{EQUITY MAIN Zone NORTH AMERICA\\In USD MEAN VAR-LOG} & {EQMAIN NAMM}  \\
\hline
 ...  & ...  \\
\hline
\end{tabular}
\captionof{table}{List of the Risk Factors for Hedge Funds Portfolio Construction}
\end{center}

we collect more than 10,000 hedge funds' data, including their monthly returns and AUMs. The selected hedge funds encompass a diverse range of strategies and characteristics. In terms of investment strategy, we have included fixed income, event driven, multi-strategy, long-short equities, macro, and various others. Geographically, the hedge funds under consideration span global, Europe, north America, Asia, and other regions. Here are some of the representatives:

\begin{center}
\begin{tabular}{||c||} 
 \hline
 Fund Name\\ [1ex] 
 \hline\hline
\makecell{400 Capital Credit Opportunities Fund LP}\\
 \hline
 \makecell{Advent Global Partners Fund}\\
 \hline
 \makecell{ Attunga Power \& Enviro Fund} \\
 \hline
 \makecell{Barington Companies Equity Partners LP}\\ 
 \hline
 \makecell{BlackRock Aletsch Fund Ltd} \\
\hline
\makecell{Campbell Managed Futures Program} \\
\hline
 ... \\
\hline
\end{tabular}
\captionof{table}{List of Hedge Funds}
\end{center}

\subsection{Benchmark description}

We select two fund of fund portfolios as the benchmarks, they are listed in Hedge Fund Research (HFR) \cite{hfr}, and let's quote their descriptions here directly:

\begin{itemize}
    \item $\bold{HFRI \  Fund \  of \  Funds \  Composite \  Index \  (HFRIFOF)}$
    \\
    
    ``Fund of Funds invest with multiple managers through funds or managed accounts. The strategy designs a diversified portfolio of managers with the objective of significantly lowering the risk (volatility) of investing with an individual manager. The Fund of Funds manager has discretion in choosing which strategies to invest in for the portfolio. A manager may allocate funds to numerous managers within a single strategy, or with numerous managers in multiple strategies. The minimum investment in a Fund of Funds may be lower than an investment in an individual hedge fund or managed account. The investor has the advantage of diversification among managers and styles with significantly less capital than investing with separate managers. PLEASE NOTE: The HFRI Fund of Funds Index is not included in the HFRI Fund Weighted Composite Index."
    
    \item $\bold{HFRI \  Fund \  Weighted \  Composite \  Index \ 
 (HFRIFWI)}$
    \\

    ``The HFRI Fund Weighted Composite Index is a global, equal-weighted index of single-manager funds that report to HFR Database. Constituent funds report monthly net of all fees performance in US Dollar and have a minimum of \$50 Million under management or \$10 Million under management and a twelve (12) month track record of active performance. The HFRI Fund Weighted Composite Index does not include Funds of Hedge Funds."
    
\end{itemize}

\subsection{Performance of the constructed portfolio}

We follow the strategy discussed in section 4 to construct our portfolios. To calculate the features based on PolyModel theory, we use the past 36 months data to compute features such as SVaR and LTS for the next month's prediction purpose. We compare the performance of our strategies against the two benchmarks from section 5.2, assuming that we start with 1 dollar at 4/30/1994; the four portfolios are SA and WA, which are based on the selection method discussed in Section 4, and the two benchmarks HFRIFOF and HFRIFWI:

\begin{center}
    \includegraphics[width=1\textwidth]{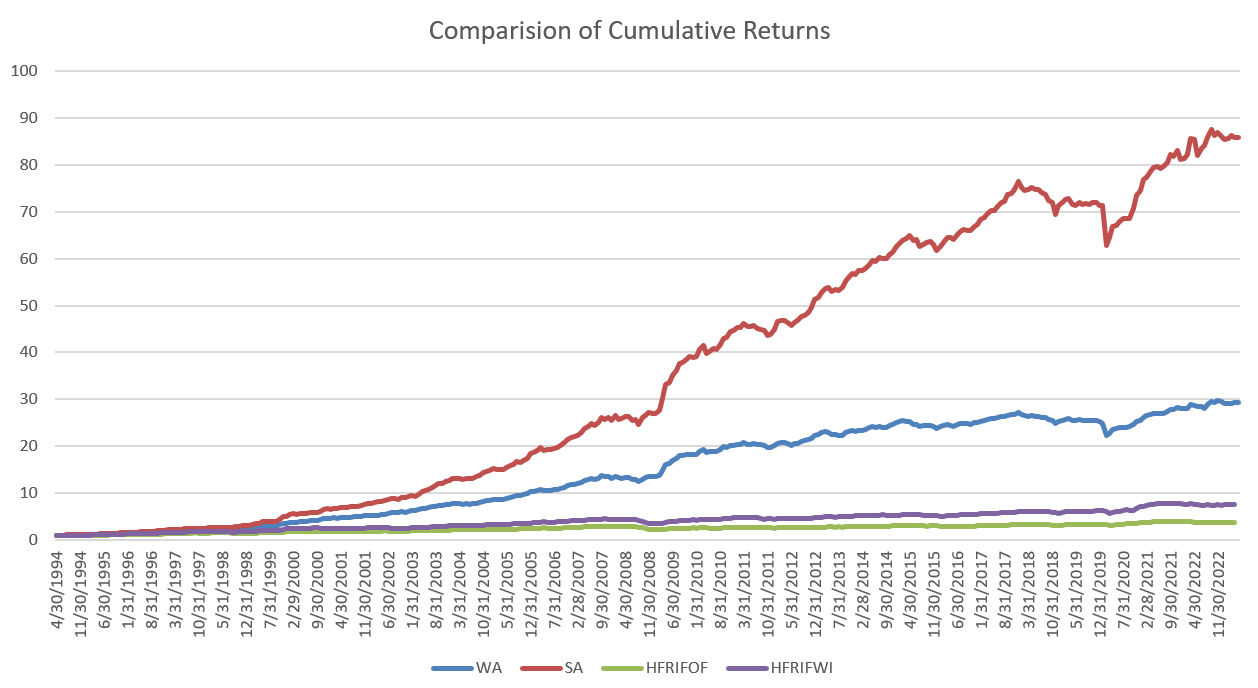}
\captionof{figure}{This figure plots the cumulative returns of the 4 strategies.}
    \label{fig:compare-cumulative-return}
\end{center}

We can see that SA has the best performance regarding the cumulative return; WA is more stable and suffers much less drawdown than SA. Both strategies outperform the benchmarks significantly. It supports the power of the combination of PolyModel feature construction and deep learning techniques.

\section{Conclusion}

In this work, we considered the problem of portfolio construction when the available data is sparse. Especially, we considered to construct a portfolio of hedge funds.

To resolve this issue, we proposed the combination of PolyModel theory and iTransformer for hedge funds selection; the proposed strategies achieved much higher returns than the standard fund of fund benchmarks. This research also shows the power of combining domain knowledge and modern deep learning techniques.

\bibliographystyle{alpha}
\bibliography{sample}

\end{document}